
\magnification=\magstep1
\def \w {\omega}

\pageno=0
{\nopagenumbers
\line{\hfil POMI Preprint -- 9/1994}
\line{\hfil November, 1994}
\vskip 3 cm
\centerline {\bf Quantum Dilogarithm as a 6j-Symbol}
\vskip 2 cm
\centerline {R.M. Kashaev}
\vskip 1 cm
\centerline{\it St. Petersburg Branch of the Steklov Mathematical
Institute,}
\centerline{\it Fontanka 27, St. Petersburg 191011, Russia}
\vskip 2 cm

\line{\bf Abstract \hfil}

The cyclic quantum dilogarithm is interpreted as a cyclic 6j-symbol
of the Weyl algebra, considered as a Borel subalgebra $BU_q(sl(2))$.
Using modified 6j-symbols, an invariant of triangulated links in triangulated
3-manifolds is constructed. Apparently,
it is an ambient isotopy invariant of links.

\vfill\eject
}
\beginsection{Introduction}

Solutions to the Yang-Baxter equation (YBE) [Y, B1] play very important role
both in the theory of solvable two-dimensional models of statistical
mechanics and field theory (for a review see [B2, F]), and in the theory
of braid groups and links [Jo]. Algebraically YBE is deeply related with
quantum groups [D1].

Recently, Rogers' dilogarithm identity [R] attracted much attention in
connection with solvable models and 3d mathematics, see [DS] and references
therein. In [FK] quantum generalization of the dilogarithm function has
been defined. Remarkably, quantized dilogarithm identity, like YBE, is also
connected with quantum groups.
In this paper we develop the result of paper [FK], and show that the quantized
Roger's (pentagon) identity for the cyclic quantum dilogarithm is a
consequence
of the co-associativity property of the Weyl algebra, endowed with particular
Hopf algebra structure (which is equivalent to Borel subalgebra of
$U_q(sl(2))$). Namely, quantum dilogarithm and cyclic 6j-symbols of the
algebra are closely related. This is the content of sections 1 and 2.
In section 3 slightly modified 6j-symbols are shown to acquire a natural
interpretation in three dimensions. In section 4 the invariant of
triangulated links in 3-manifolds is constructed.

\beginsection{1. Cyclic Representations of Weyl Algebra}

Consider unital Hopf algebra ${\cal W}$ generated by operators $E$
and $D$, satisfying Weyl permutation relations
$$
DE=\w ED              \eqno(1.1)
$$
with $\w$ being a primitive $N$ th root of unity, where odd
$N>2$, and co-multiplications [D2, Ji, FRT]
$$ \Delta(E)=E\otimes E,\quad
\Delta(D)=1\otimes D+D\otimes E.    \eqno(1.2)
$$
Elements $E^N$ and $D^N$ are central, and their
co-multiplications are
$$
\Delta(E^N)=E^N\otimes E^N,\quad
\Delta(D^N)=1\otimes D^N+D^N\otimes E^N.    \eqno(1.3)
$$
That means we can consider a restricted class of representations with
$$
E^N=1.                    \eqno(1.4)
$$
If operator $D$ is invertible, then each irreducible representation
$p$, called cyclic one, is characterized by a non-zero complex number
$z_p$, the value of central element $D^N$.
Cyclic representations of quantum groups [DK, DJMM1, AC, BK] are related
with the chiral Potts model [AMPY, MPTS, BPA] and its' generalizations
[DJMM2, BKMS].
If two representations $p$ and $q$ are such that $z_p+z_q\ne0$,
then, the tensor product of $p$ and $q$ can be
decomposed into direct sum of $N$ copies of one and the same representation
$pq$ ($pq=qp$) given by:
$$
z_{pq}=z_p+z_q.                            \eqno(1.5a)
$$
``Inverse'' representation $p^{-1}$ and ``complex conjugate'' one $p^*$
are defined by
$$
z_{p^{-1}}\equiv -z_p,\quad z_{p^*}\equiv z_p^*.        \eqno(1.5b)
$$
Introduce the ``standard'' matrix realization of representation $p$:
$$
 p(D)=x_pX, \quad p(E)=Z^{-1},    \eqno(1.6)
$$
where $N$-by-$N$ matrices $Z$ and $X$ are defined by their matrix elements
$$
\langle m|Z|n\rangle=\w^m\delta_{m,n},\quad
\langle m|X|n\rangle=\delta_{m,n+1},\quad m,n\in {\bf Z}_N,        \eqno(1.7)
$$
${\bf Z}_N$ being identified with the set $\{0,1,\ldots,N-1\}$,
and parameter $x_p$ is a particular  $N$ th root of $z_p$,
chosen to be real for real $z_p$:
$$
x_p=z_p^{1/N},\quad z_p\in{\bf R}\Rightarrow x_p\in{\bf R}.     \eqno(1.8)
$$

Let $V_p$ be an $N$ dimensional vector space where representation $p$ acts.
Clebsh-Gordon (CG) operators $K_\alpha(p,q)$, $\alpha\in {\bf Z}_N$, acting
from
$V_{pq}$ to $V_p\otimes V_q$, are linearly
independent set of solutions for the intertwining equations
$$
\eqalign{
K_\alpha(p,q)\colon &V_{pq}\to V_p\otimes V_q,\cr
&K_\alpha(p,q)pq(.)=p\otimes q\circ\Delta(.)
K_\alpha(p,q),
\quad\alpha\in {\bf Z}_N,\cr}                     \eqno(1.9)
$$
where the dot symbolizes any element of ${\cal W}$. Dual
CG operators $\overline K^\alpha(p,q)$, $\alpha\in {\bf Z}_N$,
acting from $V_p\otimes V_q$ to $V_{pq}$, are defined similarly:
$$
\eqalign{
\overline K^\alpha(p,q)\colon &V_p\otimes V_q\to V_{pq},\cr
&\overline K^\alpha(p,q)p\otimes q\circ\Delta(.)=pq(.)\overline K^\alpha(p,q),
\quad\alpha\in {\bf Z}_N.\cr}                     \eqno(1.10)
$$
Impose the following natural relations on these operators:
$$
\overline K^\alpha(p,q)
K_\beta(p,q)=\delta^\alpha_\beta{\bf1}_{pq},\quad
\sum_{\alpha\in {\bf Z}_N}K_\alpha(p,q)
\overline K^\alpha(p,q)={\bf1}_{p}\otimes{\bf1}_{q},\eqno(1.11)
$$
where, e. g. ${\bf1}_{p}$ is the identity matrix in $V_p$.

In what follows we choose  $\w^{1/2}$ as an $N$ th root of 1:
$$
\w^{1/2}=\w^{(N+1)/2}.                                       \eqno(1.12)
$$
In particular basis explicit solutions to (1.9)--(1.11) have the form:
$$
\langle
i,j|K_\alpha(p,q)|k\rangle=\langle pq,q\rangle^{1/2}
\w^{\alpha j}w(x_q,x_p,x_{pq}|i,\alpha)
\delta_{k,i+j},\quad i,j,k,\alpha\in {\bf Z}_N,       \eqno(1.13)
$$
$$
\langle
k|\overline K^\alpha(p,q)|i,j\rangle=
{\langle pq,q\rangle^{1/2}\over \langle pq\rangle}
{\delta_{k,i+j}\over\w^{\alpha j}
w(x_q/\w,x_p,x_{pq}|i,\alpha)},\quad i,j,k,\alpha\in {\bf Z}_N \eqno(1.14)
$$
where for any representations $p$ and $q$ symbol $\langle p,q\rangle$ is
defined as
$$
\langle p,q\rangle \equiv N^{-1} (x_p^N-x_q^N)/(x_p-x_q),
\quad \langle p\rangle\equiv \langle p,p\rangle=x_p^{N-1},
    \eqno(1.15)
$$
and
$$
w(x,y,z|i,j)=w(x,y,z|i-j)\w^{j^2/2},                     \eqno(1.16)
$$
$$
w(x,y,z|i)=\prod_{j=1}^i{y\over z-x\w^j}, \eqno(1.17)
$$
for complex $x,y,z$ such that
$$
x^N+y^N=z^N.                             \eqno(1.18)
$$

Using CG operators one can calculate in the standard way
6j-symbols. The defining relations are as follows:
$$
K_\alpha(p,q)K_\beta(pq,r)=\sum_{\gamma,\delta}
R(p,q,r)_{\alpha,\beta}^{\gamma,\delta}K_\delta(q,r)K_\gamma(p,qr).
\eqno(1.19)
$$
 These relations can be written in operator form, if one interprets
$R(p,q,r)_{\alpha,\beta}^{\gamma,\delta}$ as matrix elements of an linear
operator $R(p,q,r)$ in ${\bf C}^N\otimes {\bf C}^N$,
while CG operators, as operator valued vectors in ${\bf C}^N$:
$$
K_1(p,q)K_2(pq,r)=
R_{12}(p,q,r)K_2(q,r)K_1(p,qr),          \eqno(1.20)
$$
where indices $1$ and $2$ denote the two multipliers in the tensor product
space ${\bf C}^N\otimes {\bf C}^N$. The dual 6j-symbols,
$\overline R(p,q,r)_{\alpha,\beta}^{\gamma,\delta}$, are just matrix elements
of the inverse operator $R(p,q,r)^{-1}$ and they satisfy the relation:
$$
\overline R_{12}(p,q,r)K_1(p,q)K_2(pq,r)=
K_2(q,r)K_1(p,qr).                              \eqno(1.21)
$$
Relations (1.20), (1.21) are consistent, provided the 6j-symbols satisfy
the ``pentagon'' identity, which in our case looks like
$$
R_{12}(p,q,r)R_{13}(p,qr,s)R_{23}(q,r,s)=
R_{23}(pq,r,s)R_{12}(p,q,rs),                        \eqno(1.22)
$$
and in terms of dual 6j-symbols,
$$
\overline R_{23}(q,r,s)\overline R_{13}(p,qr,s)\overline R_{12}(p,q,r)=
\overline R_{12}(p,q,rs)\overline R_{23}(pq,r,s).           \eqno(1.23)
$$
Such equations were derived in [M] using geometrical arguments.
Formulas (1.13) and (1.14) lead to the following explicit expression for the
6j-symbols:
$$
R(p,q,r)_{\alpha,\beta}^{\gamma,\delta}={\rho_{p,q,r}\over\langle qr\rangle}
\w^{\alpha\delta}
w(x_{pqr}x_q,x_px_r,x_{pq}x_{qr}|\gamma,\alpha)
\delta_{\beta,\gamma+\delta},                              \eqno(1.24)
$$
where
$$
\rho_{p,q,r}={\langle pq,q\rangle^{1/2}\langle pqr,r\rangle^{1/2}
\langle qr,r\rangle^{1/2} \over \langle pqr,qr\rangle^{1/2}}
f\bigl({x_r\over x_{pqr}},{x_r\over \w x_{qr}}
|{x_{pq}x_{qr}\over x_{pqr}x_q}\bigr),                     \eqno(1.25)
$$
and function $f(x,y|z)$ is defined as
$$
f(x,y|z)=\sum_{j=0}^{N-1}{w(x|j)\over w(y|j)}z^j,\quad
w(x|i)=\prod_{j=1}^i{1\over 1-x\w^j},                        \eqno(1.26)
$$
for any $x,y,z$ satisfying the equation
$$
{1-x^N\over1-y^N}=z^N,                           \eqno(1.27)
$$
see (1.15)--(1.18) for other notations.
Function $f(x,y|z)$ has particular automorphic properties, described in [KMS2].
The dual 6j-symbols can be written in the form:
$$
\overline R(p,q,r)^{\alpha,\beta}_{\gamma,\delta}={\overline\rho_{p,q,r}\over
\langle pq\rangle}
{\delta_{\beta,\gamma+\delta}\over
\w^{\alpha\delta}
w(x_{pqr}x_q/\w,x_px_r,x_{pq}x_{qr}|\gamma,\alpha)},\eqno(1.28)
$$
 where
$$
\overline\rho_{p,q,r}={\langle pq,q\rangle^{1/2}\langle pqr,r\rangle^{1/2}
\langle qr,r\rangle^{1/2} \over \langle pqr,qr\rangle^{1/2}}
f\bigl({x_r\over x_{qr}},{x_r\over \w x_{pqr}}
|{x_{pqr}x_q\over x_{pq}x_{qr}}\bigr).                     \eqno(1.29)
$$

\beginsection{2. Quantum Dilogarithm as a 6j-Symbol}

Operator $R(p,q,r)$, given by (1.24), has the following algebraic properties:
$$
\eqalign{
R_{12}(p,q,r)Z_1Y_2&=Z_1Y_2R_{12}(p,q,r),\cr
Y_1Y_2R_{12}(p,q,r)&=R_{12}(p,q,r)Y_1,\cr
Z_2R_{12}(p,q,r)&=R_{12}(p,q,r)Z_1Z_2,\cr}                     \eqno(2.1)
$$
where
$$
Y=\w^{1/2}XZ,                                              \eqno(2.2)
$$
while $X$ and $Z$ are defined in (1.7), and subscripts denote the multipliers
in the tensor product space. Suppose we have some fixed invertible
operator $S$ in ${\bf C}^N\otimes {\bf C}^N$, also satisfying relations (2.1):
$$
S_{12}Z_1Y_2=Z_1 Y_2S_{12},\quad
Y_1 Y_2S_{12}=S_{12}Y_1,\quad
Z_2S_{12}=S_{12}Z_1 Z_2.                     \eqno(2.3)
$$
Then, combination of the form
$$
\Psi_{p,q,r}\equiv S_{12}^{-1}R_{12}(p,q,r),         \eqno(2.4)
$$
satisfies more simple relations
$$
\Psi_{p,q,r}Z_1Y_2=Z_1Y_2\Psi_{p,q,r},\quad
Y_1\Psi_{p,q,r}=\Psi_{p,q,r}Y_1,\quad
Z_1Z_2\Psi_{p,q,r}=\Psi_{p,q,r}Z_1Z_2,                     \eqno(2.5)
$$
which imply that $\Psi_{p,q,r}$ as operator is a function of a particular
combination of $Y$'s and $Z$'s:
$$
\Psi_{p,q,r}=\Psi_{p,q,r}(-Y_1^{-1}Z_2^{-1}Y_2).              \eqno(2.6)
$$
Let us choose operator $S_{12}$ as
$$
S_{12}=N^{-1}\sum_{i,j\in {\bf Z}_N}\w^{-ij}Z_1^iY_2^j.           \eqno(2.7)
$$
In addition to relations (2.3), it commutes with operator $Z_1$:
$$
S_{12}Z_1=Z_1S_{12},                                    \eqno(2.8)
$$
and satisfies the constant ``pentagon'' relation:
$$
S_{12}S_{13}S_{23}=S_{23}S_{12}.                      \eqno(2.9)
$$
We can express from (2.4) $R(p,q,r)$ in terms of operators $S$ and
$\Psi_{p,q,r}$, and substitute it into (1.22). By the use of (2.3) and (2.8)
move now all $S$'s to the left, and drop them, using (2.9). Eventually,
we end up with relation only in terms of $\Psi$'s:
$$
\Psi_{p,q,r}(U)\Psi_{p,qr,s}(-UV)\Psi_{q,r,s}(V)
=\Psi_{pq,r,s}(V)\Psi_{p,q,rs}(U),          \eqno(2.10)
$$
where
$$
U=-Y_1^{-1}Z_2^{-1}Y_2,\quad V=-Y_2^{-1}Z_3^{-1}Y_3.   \eqno(2.11)
$$
Here operators $U$ and $V$ satisfy Weyl relation and have $N$ th
powers equal to $-{\bf1}$:
$$
UV=\w VU,\quad U^N=V^N=-{\bf1}.                           \eqno(2.12)
$$
Relation (2.10) coincides with quantum dilogarithm identity (3.9) of [FK].
It is equivalent to the restricted star-triangle relation of [BB2], playing
the key role in the integrable 3d Baxter-Bazhanov model [BB1, KMS1, KMS2].
The latter is a multistate generalization of the Zamolodchikov model [Z].

\beginsection{3. Three-Dimensional Picture}

Define a matrix $T_{12}(p,q,r|a,c)$, which unlike $R_{12}(p,q,r)$ depends
on two additional ${\bf Z}_N$-arguments $a$ and $c$:
$$
T_{12}(p,q,r|a,c)\equiv \w^{ac/2}
\langle qr\rangle Y_1^{-a}Z_1^{-c}R_{12}(p,q,r)
Z_1^{c}Z_2^{-a}, \eqno(3.1)
$$
see sections 1 and 2 for the notations.
The corresponding dual matrix $\overline T_{12}(p,q,r|a,c)$ is defined
by the Hermitian conjugation of $T_{12}(p,q,r|a,c)$, combined with negation
of matrix indices and complex conjugation of continuous arguments:
$$
\overline T_{12}(p,q,r|a,c)\equiv
C_1C_2(T_{12}(p^*,q^*,r^*|a,c))^{\dagger}C_1C_2,    \eqno(3.2)
$$
where matrix $C$ is given by matrix elements of the form:
$$
C_{\alpha,\beta}=\delta_{\alpha+\beta},\quad \alpha,\beta\in {\bf Z}_N.
\eqno(3.3)
$$
Formulas (3.1) and (3.2) generalize (1.24) and (1.28)
in the sense that, if the ``charges'' $a$ and $c$ are zero, then the new
definitions coincide with (1.24) and (1.28) up to scalar factors:
$$
T_{12}(p,q,r|0,0)=\langle qr\rangle
 R_{12}(p,q,r),\quad
\overline T_{12}(p,q,r|0,0)=
\langle pq\rangle
\overline R_{12}(p,q,r).             \eqno(3.4)
$$

Define a pair of
two-index tensors $G_{\alpha,\beta}$ and $F_{\alpha,\beta}(p,q)$:
$$
G_{\alpha,\beta}=\w^{\alpha^2/2}\delta_{\alpha+\beta},\quad
\alpha,\beta\in {\bf Z}_N,                                    \eqno(3.5)
$$
and
$$
F_{\alpha,\beta}(p,q)=
{\langle p,q^{-1}p\rangle^{1/2}\over N\langle q,p^{-1}q\rangle^{1/2}}
f(0,{x_p\over\w x_{q^{-1}p}}|{x_{p^{-1}q}\over x_q})
\w^{\alpha\beta},\quad \alpha,\beta\in {\bf Z}_N.       \eqno(3.6)
$$
The corresponding inverse tensors, $G^{\alpha,\beta}$ and
$F^{\alpha,\beta}(p,q)$, are defined through the equations:
$$
\sum_{\beta}G_{\alpha,\beta}G^{\beta,\gamma}=\delta_\alpha^\gamma,\quad
\sum_{\beta}F_{\alpha,\beta}(p,q)F^{\beta,\gamma}(p,q)=
\delta_\alpha^\gamma.                                     \eqno(3.7)
$$
Using these definitions, one can prove the following formulas:
$$
\sum_{\alpha,\gamma}
T(p,q,r|a,c)_{\alpha,\beta}^{\gamma,\delta}G_{\gamma,\gamma'}
G^{\alpha,\alpha'}=
\w^{a/4}\overline
T(p^{-1},pq,r|a,1/2-a-c)^{\alpha',\delta}_{\gamma',\beta},\eqno(3.8)
$$
$$
\sum_{\alpha,\delta}
T(p,q,r|a,c)_{\alpha,\beta}^{\gamma,\delta}G_{\delta,\delta'}
F^{\alpha,\alpha'}(p,pq)=
\w^{-c/4}\overline T(pq,q^{-1},qr|1/2-a-c,c)^{\alpha',\gamma}_{\beta,\delta'}
, \eqno(3.9)
$$
$$
\sum_{\beta,\delta}
T(p,q,r|a,c)_{\alpha,\beta}^{\gamma,\delta}F_{\delta,\delta'}(q,qr)
F^{\beta,\beta'}(pq,pqr)=\w^{a/4}
\overline T(p,qr,r^{-1}|a,1/2-a-c)^{\gamma,\beta'}_{\alpha,\delta'},
                                   \eqno(3.10)
$$

A natural three-dimensional interpretation for these relations
can be given. In what follows for a polyhedron $X$, considered as a collection
of vertices, edges, triangles, and tetrahedrons,
 we will use the
standard notation $\Lambda_i(X)$, $i=0,1,2,3$ for the sets of
simplices of corresponding dimension.

 First, consider a topological \footnote{$^{\star}$}{
the term ``topological'' here means that edges and faces of the
tetrahedron can be curved}
 tetrahedron $T$ in ${\bf R}^3$.
Order the vertices by fixing the bijective map
$u\colon \{0,1,2,3\}\to\Lambda_0(T)$, $i\mapsto u_i$. Put
an arrow on each edge, pointing from a ``larger'' vertex
(with respect to above ordering) to a ``smaller'' one. The tetrahedron
 itself has two possible orientations in the following sense.
Let $u_3$ be the top of the tetrahedron.
Let us look from it down at the vertices $u_0,u_1,u_2$.
We will see two possible views: either $u_0,u_1,u_2$, in the order which they
are written, go round in the counter-clockwise direction
(the ``right'' orientation) or, in the
clockwise one (the ``left'' orientation). Introduce three maps:
$$
s\colon \Lambda_0(T)\to{\bf C},\quad c\colon \Lambda_1(T)\to{\bf Z}_N,
\quad\alpha\colon \Lambda_2(T)\to{\bf Z}_N,                 \eqno(3.11)
$$
where $s$ is injective, and $c$ satisfies the following relations:
$$
\sum_{e\in\Lambda_1(T|v)}c(e)=1/2,\quad v\in\Lambda_0(T),\quad
\Lambda_1(T|v)\equiv\{e\in\Lambda_1(T)\colon v\in e\}. \eqno(3.12)
$$
Let $c_{ij}=c(u_iu_j)$ ($u_iu_j$ is the edge having ends $u_i$ and $u_j$),
$\alpha_i=\alpha(u_ju_ku_l)$, $\{j,k,l\}=\{0,1,2,3\}\setminus\{i\}$, and
$s_{ij}$ be a representation with $z_{s_{ij}}=s(u_i)-s(u_j)$
\footnote{*}{Such a parametrization of the representations has been
suggested to the author by V.V. Bazhanov, private communication}.
Define the symbol associated with the tetrahedron $T$:  $$
T_u(s,c,\alpha)=\cases{ T(s_{01},s_{12},s_{23}|c_{01},c_{12})
_{\alpha_3,\alpha_1}^{\alpha_2,\alpha_0},& right orientation;\cr
\overline T(s_{01},s_{12},s_{23}|c_{01},c_{12})
^{\alpha_3,\alpha_1}_{\alpha_2,\alpha_0},& left orientation,\cr} \eqno(3.13)
$$

Next, consider two-sided topological triangle $F$
in ${\bf R}^3$, with one doubled edge, i.e. one pair of vertices is connected
by two different edges. As above, fix the ordering map
$u\colon \{0,1,2\}\to\Lambda_0(F)$ in such a way, that $u_0$ does not belong
to the doubled edge.
Put arrows on single edges according to the same rule as for
the tetrahedron above, while arrows on the double edges should point out to
different vertices. There are two orientations of $F$. Indeed,
if we look from the vertex
$u_0$ down at the double edges, then we will see again two possible views:
either arrows
on the double edges go round in the counter-clockwise direction
(the ``right'' orientation), or in the
clockwise one (``left'' orientation). Fix three maps:
$$
s\colon \Lambda_0(F)\to{\rm\bf C},\quad c\colon \Lambda_1(F)\to{\rm\bf Z}_N,
\quad\alpha\colon \Lambda_2(F)\to{\rm\bf Z}_N,                 \eqno(3.14)
$$
where $s$ is again injective, but $c$ is just a zero map: $c(e)=0$,
$\forall e\in\Lambda_1(F)$. Now associate with $F$ a symbol:
$$
F_u(s,c,\alpha)=\cases{
F_{\alpha_1,\alpha_2}(s_{01},s_{02}),&
right orientation;\cr
F^{\alpha_1,\alpha_2}(s_{01},s_{02}),& left orientation,\cr} \eqno(3.15)
$$
where $\{\alpha_1,\alpha_2\}=\alpha(\Lambda_2(F))$, and the other notations
are the same as in (3.13). Perform the same construction with another
two-sided topological triangle $G$ with one doubled edge, but the roles
of the vertices $u_0$ and $u_2$ being exchanged, so the double edges connect
now vertices $u_0$ and $u_1$. The corresponding symbol,
associated with $G$, has the form:
$$
G_u(s,c,\alpha)=\cases{G_{\alpha_1,\alpha_2},& right orientation;\cr
G^{\alpha_1,\alpha_2},& left orientation.\cr} \eqno(3.16)
$$

Now formulas (3.8), (3.9), and (3.10) (and their inverses) acquire the
following geometrical interpretation.
Take our topological tetrahedron $T$, choose any edge, connecting nearest
(in the sense of the defined ordering) vertices, and glue two two-sided
topological triangles to two faces, sharing this edge (the doubled edges
should match this chosen edge).
Writing this operation in terms of symbols for these geometrical objects,
one should use restrictions of the same maps $u$, $s$, and $\alpha$
in the all three symbols,
and sum over ${\bf Z}_N$-indices, corresponding to the glued faces.
The result is the
tetrahedron $T$ with the integers, corresponding to the ends of the chosen
edge, being exchanged. Namely, formula
(3.8) is described by the change of the map $u$ into $u\circ\sigma_{01}$;
(3.9), to $u\circ\sigma_{12}$;
and (3.10), to $u\circ\sigma_{23}$, where $\sigma_{ij}$ is an elementary
permutation map of integers $i$ and $j$.
These generate all tetrahedral group ${\bf S}_4$. So, we conclude that
symbol (3.13) is covariant under tetrahedral group up to $N$-th roots of
unity.

To associate three-dimensional picture with a pentagon relation, generalizing
(1.22), consider five points in ${\bf R}^3$, which are such
that any four of them are non-coplanar. There is a convex polyhedron
$W$, having these points as its vertices. Fix the ordering map
$u\colon \{0,1,2,3,4\}\to\Lambda_0(W)$.
 The symmetry group of $W$
is ${\bf S}_2\times {\bf S}_3$. Group ${\bf S}_2$ here
acts nontrivially only on two vertices of $W$, let they be
${\cal O}_2=\{u_1,u_3\}$,
while group ${\bf S}_3$
acts among remained three vertices, ${\cal O}_3=\{u_0,u_2,u_4\}$.
To fix the orientation of $W$, suppose
that, looking from vertex $u_3$, we see three vertices $u_0$, $u_2$, $u_4$
running in the clockwise direction.
$W$ can be naturally splitted either into three
tetrahedrons $T^4=u_0u_1u_2u_3$, $T^2=u_0u_1u_3u_4$ and $T^0=u_1u_2u_3u_4$,
or only into two, $T^1=u_0u_2u_3u_4$ and $T^3=u_0u_1u_2u_4$.
 The pentagon relation equates these
splittings on the level of symbols for the tetrahedrons. To make a precise
statement, fix the maps, associated with $W$,
$$
s\colon \Lambda_0(W)\to {\bf C},\quad c\colon \Lambda_1(W)\to{\bf Z}_N,\quad
\alpha\colon \Lambda_2(W)\to{\bf Z}_N, \eqno(3.17)
$$
where $s$ is injective, and $c$ satisfies the relations:
$$
\sum_{e\in\Lambda_1(W|v)}c(e)=\cases{1/2,& $v\in{\cal O}_2$;\cr
1,& $v\in{\cal O}_3$,\cr} \eqno(3.18)
$$
see also (3.12), and the individual maps, associated with the tetrahedrons:
$$
s\bigg|_{T^i}\colon \Lambda_0(T^i)\to {\bf C},\quad
c_i\colon \Lambda_1(T^i)\to {\bf Z}_N,\quad
\alpha_i\colon \Lambda_2(T^i)\to{\bf Z}_N,\quad i=0,1,2,3,4.\eqno(3.19)
$$
Here $c_i$'s satisfy restrictions (3.12) as well as
$$
c=\sum_{i=0,2,4}c_i\bigg|_{\Lambda_1(W)}=
\sum_{i=1,3}c_i\bigg|_{\Lambda_1(W)}, \eqno(3.20)
$$
where $c_i(e)=0$ if $e\not\in\Lambda_1(T^i)$; and
$$
\alpha_i\bigg|_{\Lambda_2(W)}=\alpha,\quad
\alpha_i\bigg|_{\Lambda_2(T^j)}=\alpha_j,\quad i,j=0,1,2,3,4.\eqno(3.21)
$$
The following pentagon relation holds
$$
\sum_{\alpha_0,\alpha_2,\alpha_4}\prod_{i=0,2,4}T^i_u(s,c_i,\alpha_i)
=\langle s_{13}\rangle
\sum_{\alpha_1,\alpha_3}\prod_{i=1,3}T^i_u(s,c_i,\alpha_i) \eqno(3.22)
$$
provided
$$
\sum_{i=0,2,4}c_i(u_1u_3)=1.   \eqno(3.23)
$$
Applying consequently transformations (3.8)--(3.10), and their
inverses to (3.22), we generate the whole set of
identities, corresponding to arbitrary maps $u$.

Using non-degenerateness of the symbol (3.13), we can derive other relations
having nice geometrical representations. First, define the symbol associated
with a plain triangle.
Let $D$ be a plain two-sided triangle (without doubled edges). Provide
it with the ordering map $u$ and the standard set of maps $s$, $c$, and
$\alpha$ (see the similar previous definitions above), $c$ being zero-map,
and associate the symbol:
$$
D_u(s,c,\alpha)=\delta_{\alpha_1,\alpha_2},              \eqno(3.24)
$$
where $\{\alpha_1,\alpha_2\}=\alpha(\Lambda_2(D))$.

Tetrahedrons $T^0$ and $T^2$, entering the pentagon relation (3.22),
themselves split another polyhedron $W'$ of the
same type as $W$. One can use the corresponding
pentagon identity to resplit $W'$ into three tetrahedrons, $T^1$, $T^3$, and
$\overline T^4$ (the bar denotes the opposite orientation). Multiplying now
both sides of (3.22) by inverse symbols of $T^1$ and $T^3$, we come to the
``inversion'' relation:
$$
\sum_{\alpha_0,\alpha_2} T_u(s,c,\alpha)\overline T_u(s,c',\alpha')
=\langle s_{13}\rangle\langle s_{02}\rangle\delta_{\alpha_1,\alpha_1'}
\delta_{\alpha_3,\alpha_3'}, \eqno(3.25)
$$
where $\alpha_i'=\alpha_i$ for $i=0,2$, and maps $c$ and $c'$ besides
relations (3.12) satisfy also
$$
c(u_1u_3)+c'(u_1u_3)=1,\quad c(u_0u_1)+c'(u_0u_1)=0.     \eqno(3.26)
$$
Geometrically relation (3.25) equates, up to scalar factors,
two tetrahedrons, glued along two
common faces ($u_0u_1u_3$ and $u_1u_2u_3$), to two plain triangles,
attached to each other along one edge.
Putting in (3.25) $\alpha_1=\alpha_1'$ and summing over $\alpha_1$, we get
one more relation:
$$
\sum_{\alpha_0,\alpha_1,\alpha_2} T_u(s,c,\alpha)\overline T_u(s,c',\alpha')
=N\langle s_{13}\rangle\langle s_{02}\rangle
\delta_{\alpha_3,\alpha_3'}, \eqno(3.27)
$$
where $\alpha_i'=\alpha_i$ for $i=0,1,2$, and maps $c$ and $c'$ satisfy
(3.12) and (3.26). Relation (3.27) equates, again up to scalar factors,
two tetrahedrons, glued
along three faces (which share vertex $u_3$), to a plain triangle.

\beginsection{4. Invariant of Triangulated Links}

Let $M$ be a finite triangulation of an oriented 3-dimensional
manifold without boundary. Denote by $\Lambda_i(M)$ the set of $i$-simplices
for $i=0,1,2,3$. Fix a subset of 1-simplices $L\subset\Lambda_1(M)$ in such
a way that any 0-simplex belongs to exactly two elements from $L$, so $L$
is some triangulated link in $M$, passing through the all vertices. Denote
$I=\{0,1,\ldots,K-1\}$, where $K$ is the number of vertices in $M$, and fix
the following maps:
$$
u\colon I\to\Lambda_0(M),\quad s\colon\Lambda_0(M)\to {\bf C},\quad
c_L\colon\Lambda_3(M)\times\Lambda_1(M)\to{\bf Z}_N,\quad
\alpha\colon\Lambda_2(M)\to{\bf Z}_N,   \eqno(4.1)
$$
where $u$ is bijective, $s$, injective, and $c_L$ satisfies the restrictions:
$$
e\not\in\Lambda_1(t),\quad t\in\Lambda_3(M)\Rightarrow c_L(t,e)=0;
\quad\sum_{e\in\Lambda_1(t|v)}c_L(t,e)=1/2,\quad v\in\Lambda_0(t);\eqno(4.2)
$$
$$
\sum_{t\in\Lambda_3(M)}c_L(t,e)=\cases{0,&$e\in L$;\cr
				       1,&otherwise,\cr}\eqno(4.3)
$$
see also (3.12) for the notations. Consider the following function:
$$
\langle L\rangle_M=N^{2-K}\sum_\alpha\prod_{t\in\Lambda_3(M)}
t_u(s,c_L,\alpha)\prod_{e\in\Lambda_1(M)\setminus L}\langle s(\partial e)
\rangle^{-1},\eqno(4.4)
$$
where representation $s(\partial e)$ is defined by $z_{s(\partial e)}=
s(u_j)-s(u_i)$ for $e=u_iu_j$.

\noindent
{\bf Theorem}. {\it $Q(M,L)\equiv(\langle L\rangle_M)^N$ for fixed $M$ and
$L$ depends on only $N$ and $\w$. Moreover, $Q(M,L)$ depends on only an
equivalence class of pairs $(M,L)$, which is defined as follows}.

Call two pairs $(M,L)$ and $(M',L')$ equivalent, if
one of them can be obtained from the other by a sequence of operations
(together with inverse ones), described in Section 3 and which correspond to
relations (3.22), (3.25), and (3.27)\footnote{$^{\star\star}$}{
Note that such transformations lead to singular triangulations in intermediate
steps, see [TV] for the definition}.

The proof of the theorem is straightforward and uses the results of
Section 3.

Our definition of equivalence can be considered as a simplicial analog of
the ambient isotopy of links. In this sense we have got an ambient isotopy
invariant of triangulated links. It is not clear, however, whether one can
establish bijective correspondence between the classes of triangulated links
just defined and ambient isotopy types of links. It seems plausible
that such a bijection does exist, since, elementary local transformations of
Section 3 (formulas (3.22), (3.25) and (3.27)) are the only natural ones
which one can imagine. So we formulate the following conjecture:
function $Q(M,L)$ is an ambient isotopy invariant of an unoriented link
embedded into an oriented closed 3-manifold.

Consider some examples. Assuming our conjecture to be true, we will not
concretize the used triangulations in calculations. The following formulas
are written up to $N$-th roots of unity:
$$
\langle {\rm trivial \;\;  knot}\rangle_{{\bf S}^3}=1,\eqno(4.5)
$$
$$
\langle {\rm two\;\;  unlinked\;\;  trivial\;\;  knots}\rangle_{{\bf
S}^3}=0,\quad \langle {\rm two-component\;\;  Hopf\;\;  link}\rangle_{{\bf
S}^3}=N,\eqno(4.6)
$$
$$
 \langle {\rm trefoil}\rangle_{{\bf
S}^3}=\sum_{k=0}^{N-1}(\w)_k,\quad  \langle
{\rm figure-eight\;\;  knot}\rangle_{{\bf S}^3}=
\sum_{k=0}^{N-1}|(\w)_k|^2,\eqno(4.7)
$$
where
$$
(\w)_k=\prod_{j=1}^k(1-\w^j).\eqno(4.8)
$$
Note one general property of the obtained invariant:
$$
Q(M,{\rm mirror\;\;  image\;\;  of}\;\;  L)=Q(M,L)^*, \eqno(4.9)
$$
which follows from the ``unitarity'' relation (3.2).

\beginsection{Summary}

The cyclic
quantum dilogarithm, introduced in [FK], is shown to be related
with cyclic 6j-symbols of the Weyl algebra endowed with Hopf algebra
structure (formulas (1.1), (1.2)).

The modified 6j-symbols (3.1), (3.2), (3.13) are associated with tetrahedrons
with some data on vertices, edges and faces. They differ from the usual
6j-symbols in ${\bf Z}_N$ -charges on edges of the corresponding
tetrahedron. Formulas
(3.8) -- (3.10) describe particular spatial transformations of the
tetrahedron, generating the whole tetrahedral group ${\bf S}_4$.
These symbols satisfy generalized pentagon identity (3.22) as well
as its' important consequences (3.25) and (3.27). For a pair $(M,L)$
of triangulated 3-manifold $M$ and triangulated link $L$ in it
(passing through the all vertices of $M$)
function (4.4) appears to be invariant (up to roots of unity)
under elementary moves (c.f. Alexander moves [A]), implemented by
relations (3.22), (3.25) and (3.27). Apparently, function (4.4) modulo
$N$-th roots of unity is
an ambient isotopy invariant of links. Unlike papers [TV] and [KMS], we
were not able to consider triangulated 3-manifolds themselves to construct
their invariant. The reason is the above mentioned ${\bf Z}_N$ charges.

\beginsection{Acknowledgements}

The author wishes to thank L.D. Faddeev for valuable discussions and
encouragement. It is also a pleasure to acknowledge useful comments and
suggestions from V.V. Bazhanov and V.V. Mangazeev.

\beginsection{References}

\item{[A]}
J.W. Alexander, Ann. Math. {\bf 31}, (1930) 294-322

\item{[AC]}
D. Arnaudon, A. Chakrabarti, Commun. Math. Phys. {\bf 139} (1991) 461

\item{[AMPTY]}
H. Au-Yang, B.M. McCoy, J.H.H. Perk, S. Tang, M. Yan, Phys. Lett. A{\bf123}
(1987) 219

\item{[B1]}
R.J. Baxter, Ann. Phys. {\bf70} (1972) 193

\item{[B2]}
R.J. Baxter, {\it Exactly Solved Models in Statistical Mechanics},
Academic Press, London, 1982

\item{[BB1]}
V.V. Bazhanov, R.J. Baxter, J. Stat. Phys. {\bf69} (1992) 453

\item{[BB2]}
V.V. Bazhanov, R.J. Baxter, J. Stat. Phys. {\bf71} (1993) 839

\item{[BK]}
V.V. Bazhanov, R.M. Kashaev, Commun. Math. Phys. {\bf136} (1991) 607-623

\item{[BKMS]}
V.V. Bazhanov, R.M. Kashaev, V.V. Mangazeev, Yu.G. Stroganov,
Commun. Math. Phys. {\bf138} (1991) 393

\item{[BPA]}
R.J. Baxter, J.H.H. Perk, H. Au-Yang, Phys. Lett. A{\bf128}  (1988) 138

\item{[D1]}
V.G. Drinfeld, Proc. Int. Cong. Math., Berkeley 1987, 798-820

\item{[D2]}
V.G. Drinfeld, Soviet Math. Dokl. {\bf32} (1985) 254-258

\item{[DCK]}
C. De Concini, V.G. Kac, Progr. in Math. {\bf92} (1990) 471-506

\item{[DJMM1]}
E. Date, M. Jimbo, K. Miki, T. Miwa, Preprint RIMS-703, 1990

\item{[DJMM2]}
E. Date, M. Jimbo, K. Miki, T. Miwa, Commun. Math. Phys. {\bf137} (1991) 133

\item{[DS]}
J.L. Dupont, C.H. Sah, Commun. Math. Phys. {\bf161} (1994) 265-282

\item{[F]}
L.D. Faddeev, Sov. Sci. Rev. C{\bf1} (1980) 107-155

\item{[FK]}
L.D. Faddeev and R.M. Kashaev, Mod. Phys. Lett. A, Vol. 9, No. 5 (1994)
427-434

\item{[FRT]}
L.D. Faddeev, N.Yu. Reshetikhin, L.A. Takhtajan, Algebraic Analysis, vol. 1,
Acad. Press, 1988,129-139

\item{[Ji]}
M. Jimbo, Lett. Math. Phys. {\bf10} (1985) 63-69

\item{[Jo]}
V.F.R. Jones, Ann. of Math. {\bf126} (1987) 335-388

\item{[KMS]}
M. Karowski, W. M\"uller, and R. Schrader, J. Phys. A, {\bf25}, (1992) 4847

\item{[KMS1]}
R.M. Kashaev, V.V. Mangazeev, Yu.G. Stroganov,
Int. J. Mod. Phys. A{\bf8} (1993) 587

\item{[KMS2]}
R.M. Kashaev, V.V. Mangazeev, Yu.G. Stroganov,
Int. J. Mod. Phys. A{\bf8} (1993) 1399

\item{[M]}
J.M. Maillet, Preprint ENSLAPP xxx/93, 1993

\item{[MPTS]}
B.M. McCoy, J.H.H. Perk, S. Tang, C.H. Sah, Phys. Lett. A{\bf125}
(1987) 9

\item{[R]}
L.J. Rogers, Proc. London Math. Soc. {\bf4} (1907) 169-189

\item{[TV]}
V. Turaev and O. Viro, Topology {\bf 31} (1992) 865

\item{[Y]}
C.N. Yang, Phys. Rev. Lett. (1967) 1312

\item{[Z]}
A.B. Zamolodchikov, Commun. Math. Phys. {\bf79} (1981) 489

\bye